# MODELING THE INTERCELULAR EXCHANGE OF SIGNALLING MOLECULES DEPENDING ON INTRA- AND INTER-CELLULAR ENVIRONMENTAL PARAMETERS

IGOR BALAZ and DRAGUTIN T. MIHAILOVIĆ

*Faculty of Agriculture, University of Novi Sad, Dositej Obradovic Sq. 8, Novi Sad 21000, Serbia*

Abstract - Exchange of biochemical substances is essential way in establishing communication between bacterial cells. It is noticeable that all phases of the process are heavily influenced by perturbations of either internal or external parameters. Therefore, instead to develop an accurate quantitative model of substances exchange between bacterial cells, we are interested in formalization of the basic shape of the process, and creating the appropriate strategy that allows further investigation of synchronization. Using a form of coupled difference logistic equations we investigated synchronization of substances exchange between abstract cells and its sensitivity to fluctuations of environmental parameters using methods of nonlinear dynamics.

## 1. Introduction

Communication between cells is ubiquitous in biological world. From single cell bacteria to complex eukaryotic organisms, cellular communication is a way for creating more complex structures through integration and maintaining of functioning. Organisms evolved various auxiliary ways for ensuring that transfer of signals can be performed timely and efficiently (e.g. development of vascular systems starting from early chordates). However, at the molecular level, basic scheme of signals exchange remains in the same shape: signaling molecules should reach cellular receptor, which in turn activates regulatory response, modulating production of targeted molecular species. These species then either directly or indirectly influence production of arriving signals. In this general approach, several points should be noted. Since communication is established by exchange of specific biochemical substances (substances in the further text) through surrounding environment, this process is heavily influenced by environmental factors. In single cell organisms environmental fluctuations are even more prominent since substances had to be released into external environment, which is not included into homeostasis created by the organism. Additionally, even in clonal population, and under strongly controlled environment, significant level of fluctuations of constituting parameters will remain, due to protein disorder (Dunker, et al., 2002) and so called intrinsic noise (Elowitz, et al., 2002; Swain, et al., 2002). Finally, due to thermal and conformational fluctuations, biochemical processes are inherently random (Longo & Hasty, 2006).

These facts indicate that signaling processes are able to maintain functionality despite very strong influence of both internal and external

fluctuations – a phenomenon called robustness (Barkai & Shilo, 2007; Kitano, 2007). In contrast to stability, where achieved state is maintained, here the whole functional process is in focus. Although it is one of the main aspects of functioning of living organisms, understanding of robustness is still very incomplete. Due to its very general nature it is reasonable to neglect some species-specific and molecule-specific aspects in order to investigate foundations of the robust behavior. Therefore, our focus in this paper is only on question how the oscillating system which is basically stochastic, and is inherently influenced by internal and external perturbations, can maintain its functioning? Therefore, instead to develop an accurate quantitative model of substances exchange between cells, we are rather interested for formalization of the basic shape of the process, and creating the appropriate strategy that allows further investigation of synchronization induced by fluctuations of intra- and inter- cellular environmental parameters. In Section 2, we give a short overview of general mechanism for substances exchange between two bacterial cells, representing cooperative communication process. Further, we identify main parameters of the process and derive a system of two coupled logistic equations as an appropriate model of the given process. In Section 3 we investigate synchronization of the model and its sensitivity to fluctuations of environmental parameters. Concluding remarks are given in Section 4.

## 2. Simple Model of Intercellular Exchange

### 2.1. *Empirical background*

Starting from bacteria where quorum sensing (Waters & Bassler, 2007) and colony formation (Stoodley, et al., 2002) are efficient mechanism for rapid switching between different phenotypes to sophisticated humoral control in vertebrates which ensures proper functioning of the organism as an integrated system, communication between cells is one of the main prerequisites for assembling them into the higher organized structures. Despite great variety of specific mechanisms and even greater number of molecules included, the general scheme, especially in unicellular organisms, remains fairly universal (see, for example Purves, et al. (2003)) as is seen in Figure 1, which is adopted as a scheme of intercellular exchange model we proposed in this paper.

Signaling molecules are ones which are deliberately extracted by the cell into intracellular environment, and which can affect behavior of other cells of the same or different type (species or phenotype) by means of active uptake and subsequent changes in genetic regulations. They can be excreted as either a side product of other metabolic processes, or as purposefully synthesized and transported from the cell. Once appeared in intercellular environment, they can be transported to other cells that can be affected. Let us note that the term *environment*, in this paper, comprises both (i) *intracellular environment* (inside the cell) and (ii) *intercellular environment* (that surrounds cells). Since active uptake is one of the milestones of the process, a very important factor in establishing communication is a current set of receptors and transporters in cellular membrane, during the communication process. At the same time they constitute backbone of the whole process, while simultaneously are very important source of perturbations of the process due to protein disorder and intrinsic noise. As a result, the process of exchange is constantly under inherent fluctuations of the aforementioned parameters. Another important factor is intercellular environment which could interfere with the process of exchange. It includes: distance between cells, mechanical and dynamical properties of the fluid which serves as a channel for exchange and various abiotic and biotic factors influencing physiology of the involved cells. Finally, in order to define exchange process as communication, received molecules should induce change in genetic regulations. Signaling molecules can influence production of a number of different genes but synthesis of molecules that are able to directly or indirectly affect production of arriving signals is necessity, to call this process a communication. Therefore, concentration of signaling molecules inside of the cell, that are destined to be extracted, can serve as an indicator of dynamics of the whole process of communication. These signaling molecules can be either the same for all involved cells or they can be different, acting directly or indirectly on production of arriving signals.

Additionally, the influence of affinity in functioning of living systems is also an important issue. It can be divided into following aspects: (a1) affinity of genetic regulators towards arriving signals which determine intensity of cellular response and (a2) affinity for uptake of signaling molecules. First aspect is genetically determined and therefore species specific. Second aspect is more complex and is influenced by: affinity of receptors to binding specific signaling molecule, number of active receptor and their conformational fluctuations (protein disorder).

**2.2. *Model philosophy***

As it is obvious from the empirical description, we can infer successfulness of the communication process by monitoring: (i) number of signaling molecules, both inside and outside of the cell and (ii) their mutual influence. Concentration of signaling molecules in intercellular environment is subject to various environmental influences, and taken alone often can indicate more about state of the environment then about the communication itself. Therefore, we choose to follow concentration of signaling molecules inside of the cell as the main indicator of the process. In that case, parameters of the system are: (i) affinity $p$ by which cells perform uptake of signaling molecules (a2), that depends on number and state of appropriate receptors, (ii) concentration $c$ of signaling molecules in intercellular environment within the radius of interaction, (iii) intensity of cellular response (a1) $x_n$ and $y_n$ and (iv) influence of other environmental factors which can interfere with the process of communication. In this case we postulate parameter $r$, that can be taken collectively for intra- and inter- cellular environment, inside of the one variable, indicating overall disposition of the environment to the communication process.

The time development ($n$ is the number of time step) of the concentration in cells $(x_n, y_n)$ can be expressed as

$$x_{n+1} = (1-c)\Psi(x_n) + h\ (\Psi(y_n)), \qquad (1a)$$

$$y_{n+1} = (1-c)\Psi(x_n) + h\ (\Psi(x_n)). \qquad (1b)$$

The map, $h$ represents the flow of materials from cell to cell, and $h(x)$ and $h(y)$ are defined by a map that can be approximated by a power map,

$$h(x) \sim cx^p, \qquad\qquad (2a)$$

$$h(y) \sim cy^q. \qquad\qquad (2b)$$

If $h(x) \sim cx^p$ and $h(y) \sim cy^q$, the interaction is expressed as a nonlinear coupling between two cells. The dynamics of intracellular behavior is expressed as a logistic map (e.g., (Deverney, 1986; Gunji & Kamiura, 2004)),

$$\Psi(x_n) = r\, x_n(1 - x_n), \qquad\qquad (3a)$$

$$\Psi(y_n) = r\, y_n(1 - y_n). \qquad\qquad (3b)$$

Since concentration of signaling molecules can be regarded as their population for fixed volume, and since we are focused on mutual influence of these populations, it points out to use the coupled logistic equations. Instead of considering cell-to-cell coupling of two explicit n-gene oscillators (Ullner, et al., 2008) we consider generalized case of gene oscillators coupling. In that case investigation of conditions under which two equations are synchronized and how this synchronization behaves under changes of intra- and inter-cellular environment, can give some answers on the question of maintaining functionality in the system. Therefore, having in mind that (i) cellular events are discrete (Barkay & Shilo, 2007) and (ii) the aforementioned reasoning, we consider system of difference equations of the form

$$X_{n+1} = F(X_n) \equiv L(X_n) + P(X_n), \qquad (4)$$

with notation

$$L(X_n) = ((1-c)rx_n(1-x_n),(1-c)ry_n(1-y_n)), \qquad P(X_n) = (cy_n^p, cx_n^{1-p}), \quad (5)$$

where $X_n = (x_n, y_n)$ is a vector representing concentration of signaling molecules inside of the cell, while $P(X_n)$ denotes stimulative coupling influence of members of the system which is here restricted only to positive numbers in the interval (0,1). The starting point $X_0$ is determined so that $(x_0, y_0) \in (0,1)$. Parameter $r \in (0,4)$ is so-called logistic parameter, which in logistic difference equation determines an overall disposition of the environment to the given population of signaling molecules and exchange processes. Affinity to uptake signaling molecules is indicated by $p$. Let us note that we require that sum of all affinities of cells $p_i$ exchanging substances has to satisfy condition $\sum_i p_i = 1$ or in the case of two cells $p + q = 1$. Since fixed point is $F(0) = 0$, in order to ensure that zero is not at the same time the point of attraction, we defined $p \in (0,1)$ as an exponent. Finally, $c$ represents coupling of two factors: concentration of signaling molecules in intracellular environment and intensity of response they can provoke. This form is taken because the effect of the same intracellular concentration of signaling molecules can vary greatly with variation of affinity of genetic regulators for that signal, which is further reflected on the ability to synchronize with other cells. Therefore, $c$ influence both, rate of intracellular synthesis of signaling molecules, as well as synchronization of signaling processes between two cells, so the parameter $c$ is taken to be a part of both $L(X_n)$ and $P(X_n)$. However, relative ratio of these two influences depends on current model setting. For example, if for both cells $X_n$ is strongly influenced by intracellular concentration of signals, while they can provoke relatively smaller response then the form of equation will be

$$x_{n+1} = (1-c)rx_n(1-x_n) + cy_n^p, \qquad (6a)$$

$$y_{n+1} = (1-c)ry_n(1-y_n) + cx_n^{1-p}. \qquad (6b)$$

## 3. Analysis of the Coupled Maps Representing the Intercellular Exchange of Substances Using Methods of Non Linear Dynamics

In order to further investigate the behavior of the coupled maps, we perform a numerical analysis of the coupled system (6) throuhg its parameters $c$, $r$ and $p$, using the largest Lyapunov exponet and cross sample entropy as measures of the chaotic behaviour and border between synchronized and nonsynchronized system states in intercellular exchange of substances.

### 3.1. *Lyapunov exponent of the coupled maps (6) for* $r = const.$

We calculate Lyapunov exponent by analysis of orbits. The orbit of the point $X_0$ is the sequence $X_0, F(X_0), ..., F^n(X_0), ...$ where $F^0(X_0) \equiv X_o$ and for $n \geq 1$, $F^n(X_0) = F(F^{n-1}(X_0))$. We say that the orbit is periodic with period $k$ if $k$ is the smallest natural number such that $F^k(X_0) = X_0$. If $k = 1$, then the point $X_0$ is the fixed point. The periodic point $X_0$ with period $k$ is an attraction point if the norm of the Jacobi matrix for the mapping $F^k(X) = (f_k(x, y)), (g_k(x, y))$ is less than one, i.e., $\| J^k(X_0) \| < 1$, where

$$J^k(X_0) = \begin{bmatrix} \dfrac{\partial f_k}{\partial x} & \dfrac{\partial f_k}{\partial y} \\[2mm] \dfrac{\partial g_k}{\partial x} & \dfrac{\partial g_k}{\partial y} \end{bmatrix}_{X = X_0} . \qquad (7)$$

Here, we define $\| J^k(X_0) \|$ as max $\{| \lambda_1 |, | \lambda_2 |\}$, where $\lambda_1$ and $\lambda_2$ are the eigenvalues of the matrix. In order to characterize the asymptotic behavior of the orbits, we need to calculate the largest Lyapunov exponent, which is given for the initial point $X_0$ in the attracting region by

$$\lambda = \lim_{n \to \infty} (\ln \| J^n(X_0) \| / n) . \qquad (8)$$

With this exponent, we measure how rapidly two nearby orbits in an attracting region converge or diverge. In practice, we compute the approximate value of $\lambda$ by substituting in (8) successive values from $X_{n_0}$ to $X_{n_1}$, for $n_0, n_1$ large enough to eliminate transient behaviors and provide good approximation. If $X_0$ is part of a stable periodic orbit of period $k$, then $\| J^k(X_0) \| < 1$ and the exponent $\lambda$ is negative, which characterizes the rate at which small perturbations from the fixed cycle decay, and we can call such a system synchronized one.

We considered a two-cell system, where each of them is able to release and uptake the same substance. According to the assumption in model design, the dynamical behavior of the substance concentrations $x_n$ and $y_n$ depends on three factors: (i) its own concentration $c$ within radius of interaction in surrounding environment, (ii) parameter $r$ and (iii) affinity $p$ for binding on cellular receptors. First factor is determined by underlying feedback mechanism of intracellular regulations, while the second one represents level of the sutiability of the environment to the communication between two cells (Mihailović, et al., 2010). The third factor depends on protein disorder (Dunker, et al., 2002). The variation of Lyapunov exponent $\lambda$ as a function of concentration $c$ is depicted in Figure 2 for $p = 0.5$ and $r = 3.95$.

İt is seen when values of $c$ excides values of 0.4 then complete synchronization (Lyapunov exponent is less than zero) in intercellular exchange of substances is achieved. In contrast to that, for values of $c$ smaller than 0.4 there exists region of non synchronized states in exchange with some windows where exchange of substances between two cells is synchronized.

In this subsection we further consider the behavior of coupling, and estimate how a coupled map system can achieve synchronization in intercellular exchange of substances depending on parameter $c$ (concentration), for a fixed value of $r$ (in our case 3.95) and different values of affinity $p$. In that purpose we calculate Lyapunov exponent of the coupled maps, given as a function of the coupling parameter $c$ ranging from 0 to 1.0, for different values of the affinity $p$ as it depicted in Figure 3 where Lyapunov exponets are calculated for $p = 0.4$, 0.3, 0.2, 0.1 and 0.0. If we look at all panels it is seen that there is an border in values of concentration $c$ (around 0.4), that split domain of concentration into two regions. The first one, tha is located between 0 and 0.4, with the non sinchronyzed states including sporadical windows where synchronization is reached. In contrast to that, the second region (between 0.4 and 1.0) is region where process of ehcange between two cells is fully synchronized. Because of the symmetry of the coupled system (6), the same results will be obtained for values $p = 0.6$, 0.7, 0.8, 0.9 and 1.0 corresponding to those for $p = 0.4$, 0.3, 0.2, 0.1 and 0.0.

### 3.2. *Entropy of the system of the coupled maps (6)*

Estimation of the system of coupled maps (6) complexity, through analysis of concentration in cells $(x_n, y_n)$ depending on on intra- and inter- cellular parameters, is of great interest for modelling procedure. In this paper, we use the sample entropy (SampEn) as a measure of the complexity of the system considered. Sample entropy, a measure quantifying regularity and complexity, is believed to be an effective analysing method of diverse settings that include both deterministic chaotic and stochastic processes, particularly operative in the analysis of physiological, sound, climate and environmental interface or cell signals that involve relatively small amount of data (Pincus, 1991; Richman & Moorman, 2000). Practically, we consider cross sample entropy $(Cross - SampEn)$ - measure of asynchrony recently introduced technique for comparing two different time series to assess their degree of asynchrony or dissimilarity (Kennel, et al., 1992; Richman & Moorman, 2000; Lake, et al., 2002). Let $u = [u(1), u(2), \ldots u(N)]$ and $v = [v(1), v(2), \ldots v(N)]$ fix input parameters $m$ and $\rho$. Vector sequences: $x(i) = [u(i), u(i+1), \ldots u(i+m-1)]$ and $y(j) = [v(j), v(j+1), \ldots v(j+m-1)]$ while $N$ is the number of data points of time series, $i, j = N - m + 1$. For each $i \le N - m$ set $B_i^m(\rho)(v \| u) = $ (number of $j \le N - m$ such that $d[x_m(i), y_m(j)] \le \rho]$) $/(N-m)$, where $j$ ranges from 1 to $N - m$.

And then

$$B^m(\rho)(v \| u) = \sum_{i=1}^{N-m} B_i^m(\rho)(v \| u) / N - m \qquad (9)$$

which is the average value of $B_i^m(v \| u)$. Similarly we define $A^m$ and $A_i^m$ as $A_i^m(\rho)(v \| u) = $ (number of such $j \le N - m$ that $d[x_m(i), y_m(j)] \le \rho]$) $/(N-m)$.

$$A^m(\rho)(v \| u) = \sum_{i=1}^{N-m} A_i^m(\rho)(v \| u) / N - m \qquad (10)$$

which is the average value of $A_i^m(v \| u)$. And then

$$Cross - SampEn\,(m, \rho, n) = -\ln\left\{ A^m(\rho)(v \| u) / B^m(\rho)(v \| u) \right\} \quad (11)$$

We applied $Cross - SampEn$ with $m = 5$ and $\rho = 0.05$ for $x_n$ and $y_n$ time series.

Figure 4 depicts cross sample entropy of the coupled maps, given as a function of the coupling parameter $c$ ranging from 0.0 to 1.0, for value of the affinity $p = 0.5$ and r = 3.95. It is seen a high disorder in the system up to the concentration $c = 0.4$. After that value there is a complete synchronization in the substances excahge. Similar behavior we obatin for different values of affinity $p$ (Figure 5). These data are in agreement with analysis of Lyapunov exponent performed in section 3.1, which indicate compatibility of used measures.

## 4. Conclusions

In this paper, our focus is on modeling synchronization in intercellular exchange of substances. We gave a short overview of general mechanism for substances exchange between two cells, representing cooperative communication process. We identified main parameters of the process and derived a system of two coupled logistic equations as an appropriate model of the given process. Then we investigated synchronization of the model and its sensitivity to fluctuations of environmental parameters using methods of nonlinear dynamics, i.e. the largest Lyapunov exponent and cross sample entropy as measures. Results show that both measures are compatible and can be used interchangeably. Both of them show existence of stability regions where noise in the form of fluctuations in concentration of signaling molecules in intercellular environment and fluctuations in affinity for uptake these molecules cannot interfere with the process of exchange. Since our model is insipred by the general scheme of intercellular communication, it naturally does not allow detailed modelling of some concrete, emiprically verifable intercellular communication process. Instead, it is designed to serve as a starting tool in general investigation of robustness in mutually stimulative populations which can be readily extended to investigation of synchronization in larger networks of interacting entities (Amritkar & Jalan, 2003; Jalan, et al., 2005).

**Acknowledgments.** The research work described here has been funded by the Serbian Ministry of Science and Technology under the project "Study of climate change impact on environment: Monitoring of impact, adaptation and moderation", for 2011-2014.

# МОДЕЛОВАЊЕ МЕЂУЋЕЛИЈСКЕ РАЗМЕНЕ СИГНАЛНИХ МОЛЕКУЛА У ЗАВИСНОСТИ ОД ИНТРА- И ИНТЕРЋЕЛИЈСКИХ ПАРАМЕТАРА

ИГОР БАЛАЖ и ДРАГУТИН Т. МИХАИЛОВИЋ

*Пољопривредни факултет,Универзитет у Новом Саду, Трг Доситеја Обрадовића 8,21000 Нови Сад, Србија*

Размена биохемијских супстанци је основни начин успостављања комуникације између бактеријских ћелија. Специфично је то да су све фазе процеса размене под значајним утицајем флуктуација спољашњих и унутрашњих параметара. Стога, уместо развијања детаљног квантитативног модела размене супстанци између бактеријских ћелија, наш циљ је формализовање основног оквира процеса и креирање одговарајуће стратегије која омогућава даље истраживање синхронизације код живих система. Користећи специфичну форму спрегнутих диференцних логистичких једначина, као и методе нелинеарне динамике, извели смо анализу синхронизације размене супстанци између апстрактних ћелија као и испитивање осетљивости на флуктуирање спољашњих параметара.

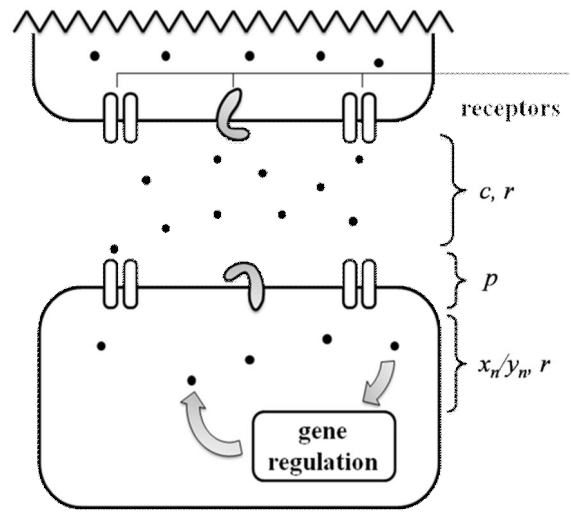

**Figure 1.** Schematic representation of intercellular communication. Here, $c$ represents concentration of signaling molecule in intercellular environment coupled with intensity of response they can provoke while $r$ includes collective influence of environment factors which can interfere with the process of communication. $x_n$ and $y_n$ represent concentration of signaling molecules in cells environment, while $p$ denotes cellular affinity to uptake the substances.

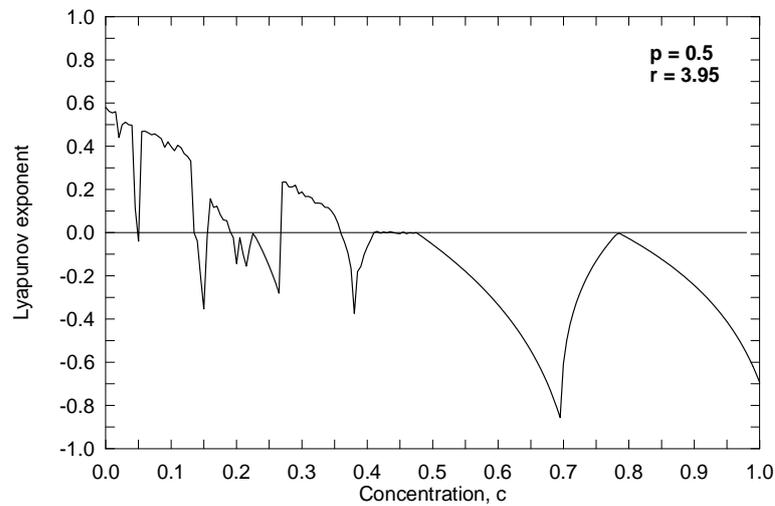

**Figure 2.** Lyapunov exponent of the coupled maps, given as a function of the coupling parameter $c$ ranging from 0.0 to 1.0, for value of the affinity $p = 0.5$. Each point in the above graphs was obtained by iterating many times (2000 iterations) from the initial condition to eliminate transient behavior and then averaging over another 500 iterations. Initial condition: $x = 0.3$, $y = 0.5$, with $200\,c$ values.

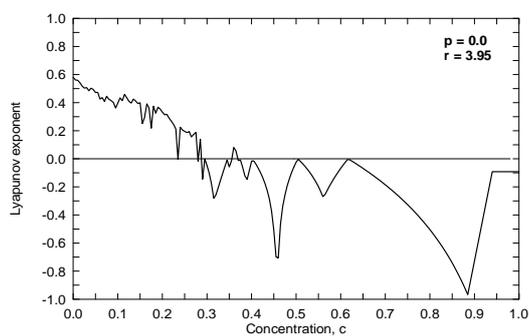

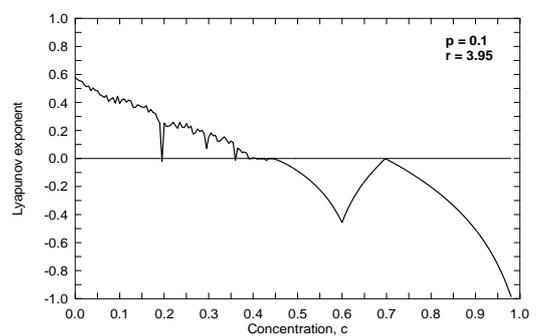

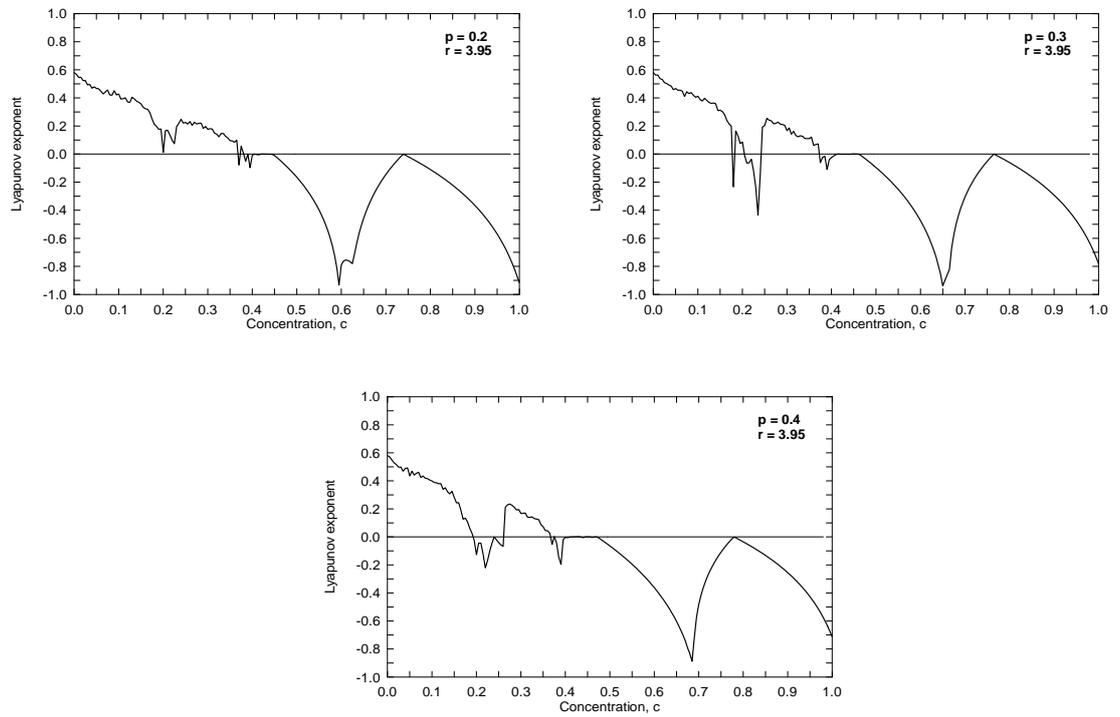

**Figure 3.** Lyapunov exponent of the coupled maps, given as a function of the coupling parameter $c$ ranging from 0.0 to 1.0, for different values of the affinity $p$. The same graphs will be able to obtained for values $p = 0.6$, 0.7, 0.8, 0.9 and 1.0 corresponding to those for $p = 0.4$, 0.3, 0.2, 0.1 and 0.0. Each point in the above graphs was obtained by iterating many times (2000) from the initial condition to eliminate transient behavior and then averaging over another 500 iterations. Initial condition: $x = 0.3$, $y = 0.5$, with $200c$ values.

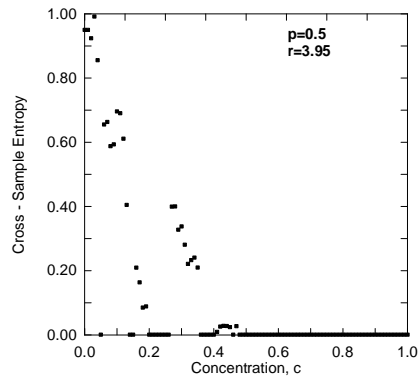

**Figure 4.** Cross sample entropy of the coupled maps, given as a function of the coupling parameter *c* ranging from 0.0 to 1.0, for value of the affinity *p* = 0.5. The $x_n$ and $y_n$ time series in the above graphs was obtained by iterating many times (2000 iterations) from the initial condition to eliminate transient behavior and then averaging over another 2000 iterations. Initial condition: *x* = 0.3, *y* = 0.5, with 200*c* values.

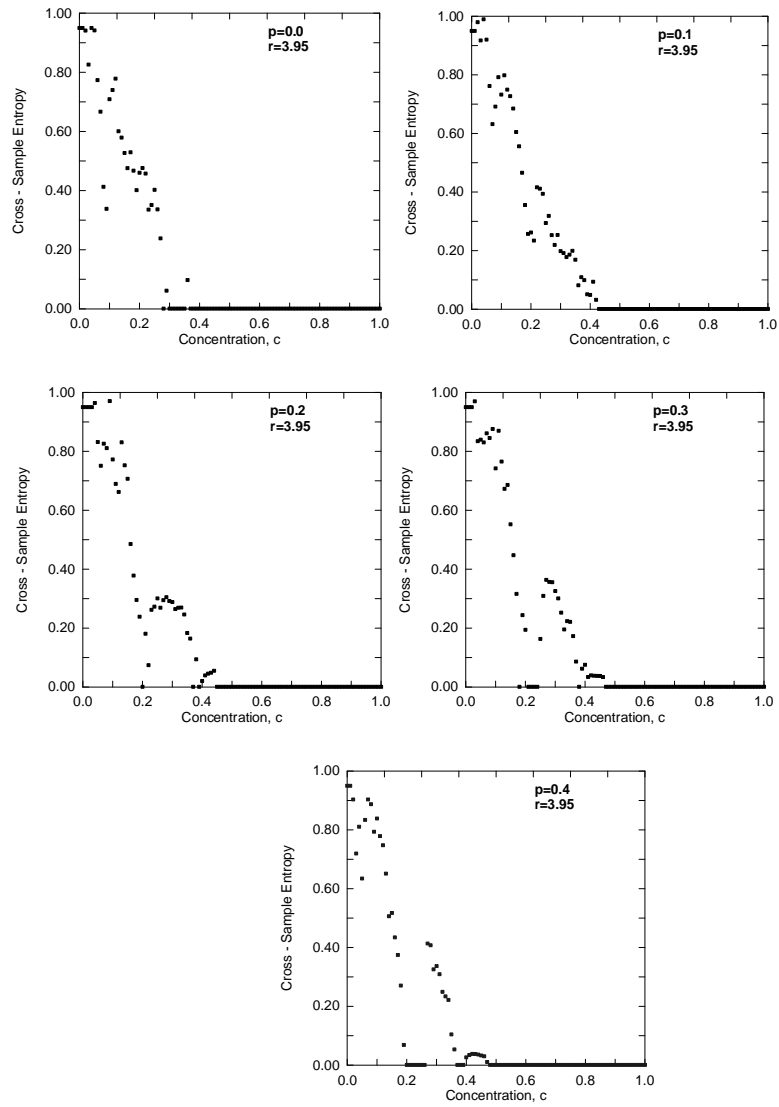

**Figure 5.** Cross sample entropy of the coupled maps, given as a function of the coupling parameter $c$ (concentration) ranging from 0.0 to 1.0, for different values of the affinity $p$. The same graphs will be able to obtained for values $p$ = 0.6, 0.7, 0.8, 0.9 and 1.0 corresponding to those for $p$ = 0.4, 0.3, 0.2, 0.1 and 0.0. The $x_n$ and $y_n$ time series in the above graphs were obtained by iterating many times (2000 iterations) from the initial condition to eliminate transient behavior and then averaging over another 2000 iterations. Initial condition: $x$ = 0.3, $y$ = 0.5, with 200$c$ values.